\documentclass[twocolumn,epjc3]{svjour3}          

\RequirePackage[T1]{fontenc}

\smartqed  
\RequirePackage{graphicx}
\RequirePackage{mathptmx}      
\RequirePackage{flushend}
\RequirePackage{upgreek}
\RequirePackage[numbers,sort&compress]{natbib}
\RequirePackage[colorlinks,citecolor=blue,urlcolor=blue,linkcolor=blue]{hyperref}

\newcommand{\de}{\mathrm{de}}
\newcommand{\dm}{\mathrm{dm}}

\begin{document}

\title{Interacting dark energy in the dark $SU(2)_R$ model}


\author{Ricardo G. Landim\thanksref{e1,addr1, addr2}
        \and
        Rafael J. F. Marcondes \thanksref{e2,addr1} 
        \and 
        Fabr\'izio F. Bernardi\thanksref{e3,addr1}
        \and 
        Elcio Abdalla\thanksref{e4,addr1}
}

\thankstext{e1}{rlandim@if.usp.br}
\thankstext{e2}{rafaelmarcondes@usp.br}
\thankstext{e3}{bernardiff@gmail.com}
\thankstext{e4}{eabdalla@if.usp.br}

\institute{Instituto de F\'isica, Universidade de S\~ao Paulo\\
 Rua do Mat\~ao,
1371, Butant\~a, CEP 05508-090, S\~ao Paulo, SP, Brazil\label{addr1}
          \and
        Departamento de F\'isica e Astronomia,\\
Faculdade de Ci\^encias da Universidade do Porto,\\
Rua do Campo Alegre 687, 4169-007, Porto, Portugal\label{addr2}
}

\date{Received: date / Accepted: date}

\maketitle

\begin{abstract}
We explore the cosmological implications of the interactions among the dark
particles in the dark $SU(2)_R$ model. It turns out that the relevant
interaction is between dark energy and dark matter, through a decay process.
With respect to the standard $\Lambda$CDM model,
it changes only the background equations. 
We note that the observational aspects of the model are dominated by
degeneracies between the parameters that describe the process. Thus, only the usual $\Lambda$CDM parameters, such as the Hubble expansion rate and the dark
energy density parameter (interpreted as the combination of the densities of the
dark energy doublet) could be constrained by observations at this moment.
\end{abstract}

\section{Introduction}

Understanding the nature of  dark energy and dark matter is a puzzling challenge
that has motivated physicists to develop huge observational programs. This is
one of the biggest concerns in modern cosmology. The simplest dark energy
candidate is a cosmological constant, in agreement with the Planck satellite
results \cite{Planck2013cosmological}. Such attempt, however,  suffers from a
huge discrepancy of 120 orders of magnitude between a theoretical prediction and
the observed data \cite{Weinberg:1988cp}. The origin of such a constant is still
an open issue which motivates physicists to look into  more sophisticated
models. The plethora of dark energy candidates  include  scalar fields
\cite{peebles1988,ratra1988,Frieman1992,Frieman1995,Caldwell:1997ii,Padmanabhan:2002cp,Bagla:2002yn,ArmendarizPicon:2000dh,Brax1999,Copeland2000,Landim:2015upa,micheletti2009},
vector fields
\cite{Koivisto:2008xf,Bamba:2008ja,Emelyanov:2011ze,Emelyanov:2011wn,Emelyanov:2011kn,Kouwn:2015cdw,Landim:2016dxh},
holographic dark energy
\cite{Hsu:2004ri,Li:2004rb,Pavon:2005yx,Wang:2005jx,Wang:2005pk,Wang:2005ph,Wang:2007ak,Landim:2015hqa,Li:2009bn,Li:2009zs,Li:2011sd,Saridakis:2017rdo,Mamon:2017crm,Mukherjee:2017oom,Mukherjee:2016lor,Feng:2016djj,Herrera:2016uci,Forte:2016ben},
models of false vacuum decay
\cite{Szydlowski:2017wlv,Stachowski:2016zpq,Stojkovic:2007dw,Greenwood:2008qp,Abdalla:2012ug,Shafieloo:2016bpk},
modifications of gravity and different kinds of cosmological fluids
\cite{copeland2006dynamics, dvali2000, yin2005}.  In addition, the two
components of the dark sector may interact with each other
\cite{Wetterich:1994bg,Amendola:1999er,Guo:2004vg,Cai:2004dk,Guo:2004xx,Bi:2004ns,Gumjudpai:2005ry,yin2005,Wang:2005jx,Wang:2005pk,Wang:2005ph,Wang:2007ak,Costa:2013sva,Abdalla:2014cla,Costa:2014pba,Costa:2016tpb,Marcondes:2016reb,Landim:2016gpz,Wang:2016lxa,Farrar:2003uw,Abdalla:2012ug,micheletti2009,Santos:2017bqm,Yang:2017yme,Marttens:2016cba,Yang:2017zjs,Xu:2017rfo},
since their densities are comparable and the interaction can eventually
alleviate the coincidence problem \cite{Zimdahl:2001ar,Chimento:2003iea}.  

Much closer to the Standard Model (SM) of particle physics are models based on
gauge groups which aim to take dark matter into account. Those with
$SU(2)_R$ symmetry, for instance, are known in the literature as extensions of
the SM, in the so-called left-right symmetric models
\cite{Hewett:1988xc,Aulakh:1998nn,Duka:1999uc,Dobrescu:2015qna,Dobrescu:2015jvn,Ko:2015uma,Brehmer:2015cia},
where, recently, dark matter has been considered as well
\cite{Bezrukov:2009th,Esteves:2011gk,An:2011uq,Nemevsek:2012cd,Bhattacharya:2013nya,Heeck:2015qra,Garcia-Cely:2015quu,Berlin:2016eem,Dev:2016qeb,Dev:2016qbd,Dev:2016xcp},
but with no attempt to insert dark energy in it.

The dark $SU(2)_R$ model was built to have the two elements of the dark sector
\cite{Landim:2016isc} and it is similar to the well-known model of weak
interactions. In principle, the hidden sector interacts with the SM particles
only through gravity. Dark energy is interpreted as a scalar field whose
potential is a sum of even self-interactions up to order six. The scalar field
is at the local minimum of the potential, and such false vacuum might decay into
the true one through the barrier penetration. However, in order to explain the
current cosmic acceleration, the false vacuum should be long-lived (with a life time of the order of the
age of the universe, as shown in \cite{Landim:2016isc})  and therefore the
scalar field behaves as a cosmological constant. On the other hand, it differs
from the latter due to the presence of interactions among the dark particles. 

In this work we explore the interactions among the dark particles from the 
cosmological point of view. The relevant interaction is among dark energy and
dark matter, through the decay process calculated in \cite{Landim:2016isc}. It
turns out that the coupling changes only the background equations, since the dark
energy perturbation decreases faster than radiation. 
The paper is organized in the following manner. Sect.~\ref{DSU2model} presents
the dark $SU(2)_R$ model, introduced in \cite{Landim:2016isc}. In
Sect.~\ref{modelpred} we derive its cosmological equations and discuss the
outcome of confronting it with observational data from the standard cosmic
probes. Sect.~\ref{conclusion} is reserved for conclusions.

\section{The dark $SU(2)_R$ model}\label{DSU2model}

In the dark $SU(2)_R$ model \cite{Landim:2016isc}, dark energy and dark matter
are doublets under $SU(2)_R$ and singlets under any other symmetry. The model
contains a  dark matter candidate $\psi$,  a dark matter neutrino $\nu_d$ (which
can be much lighter than $\psi$), and the dark energy doublet $\varphi$, with
$\varphi^0$ and $\varphi^+$,  the latter being  the heaviest particle by
definition. The scalar potential for the dark energy doublet is
\begin{equation}\label{VScalar}
 V(\Phi)=\frac{m^2}{2}\Phi^{\dagger}\Phi-\frac{\lambda}{4}(\Phi^{\dagger}\Phi)^2+\frac{g'}{\Lambda^2}(\Phi^{\dagger}\Phi)^3 \quad,
\end{equation} 
where $\lambda$ and $g'$ are positive constants and $\Lambda$ is the cutoff
scale. There are also terms which involve couplings with the dark Higgs so that
after the spontaneous symmetry breaking the physical mass of the dark energy
doublet is no longer the same, but $m_{\varphi^0}$ and  $m_{\varphi^+}$.

The dimension six interaction term can be split into 
\begin{equation}\label{VScalar6}
\frac{g'}{\Lambda^2}(\Phi^{\dagger}\Phi)^3=\left[\frac{\lambda}{32 m_i^2}+\frac{g}{\Lambda^2}\right](\Phi^{\dagger}\Phi)^3 \quad,
\end{equation}
where $g'= \frac{\lambda^2 }{32 m_i^2}\Lambda^2-g$ and $i$ stands for $\varphi^0$ or $\varphi^+$. 

The mass term, the quartic interaction and the first term of the dimension six
operator can be grouped as a perfect square,
$\bigl[\frac{m_i \varphi^i }{ \sqrt{2}}-\frac{\lambda (\varphi^{i})^3}{\sqrt{32} \, m_i} \bigr]^2$,
which has an absolute minimum at $V(\varphi^i)=0$. The extra term
$g\Lambda^{-2}(\varphi^i)^6$ brings the minimum  $V(\pm 2m/\sqrt{\lambda})$
downwards, thus the difference between the true vacuum and the false one, given
by $V(0)-V(\pm 2\,m_i/\sqrt{\lambda})= 64 \, m_i^6\lambda^{-3} g\Lambda^{-2}$, is the
observed vacuum energy. A gravity-induced term ($g M_{pl}^{-2}(\varphi^{i})^6$), which may  parametrizes  a graviton loop contribution \cite{ArkaniHamed:2008ym},
is a natural option since we are dealing with gravitational effects,  therefore,
the  reduced Planck mass is the cutoff scale.\footnote{It is possible to get the
    scalar potential (\ref{VScalar}) from minimal supergravity, for instance.
    However, as usual in supergravity theories, we ended up with a negative
    cosmological constant.}  The mass of the scalar field should be, for instance, $\sim
\mathcal{O}$(GeV) for $\lambda\sim g \sim 1$ in order to explain the observed
value of $10^{-47}$~GeV$^4$. The value of the observed vacuum energy constrains one of the three parameters, namely, $m_i$, $\lambda$, or $g$.

The interaction between the fields are given by the Lagrangian
\begin{eqnarray}
 \mathcal{L}_{int}=
g\left(W_{d\mu}^+J_{dW}^{+\mu} +W_{d\mu}^-J_{dW}^{-\mu}+Z_{d\mu}^0J_{dZ}^{0\mu}\right)\quad,
\label{eq:Lint}
\end{eqnarray}
where the currents are 
\begin{eqnarray}
 &J_{dW}^{+\mu} =\frac{1}{\sqrt{2}}[\bar{\nu}_{dR}\gamma^\mu \psi_R+i(\varphi^0\partial^\mu\bar{\varphi}^+-\bar{\varphi}^+\partial^\mu\varphi^0)]\quad , \label{eq:JW+}\\
& J_{dW}^{-\mu} = \frac{1}{\sqrt{2}}[\bar{\psi}_{R}\gamma^\mu \nu_{dR}+i(\varphi^+\partial^\mu\bar{\varphi}^0-\bar{\varphi}^0\partial^\mu\varphi^+)]
\quad ,\label{eq:JW-}\\
&J_{dZ}^{0\mu} =\frac{1}{2}[\bar{\nu}_{dR}\gamma^\mu \nu_{dR}-\bar{\psi}_{R}\gamma^\mu \psi_{R}+
i(\varphi^+\partial^\mu\bar{\varphi}^+-\bar{\varphi}^+\partial^\mu\varphi^+)\nonumber\\ & -i(\varphi^0\partial^\mu\bar{\varphi}^0-\bar{\varphi}^0\partial^\mu\varphi^0)]\quad .
\label{eq:JZ0}
\end{eqnarray}

The behaviour of this system is two fold. If we place the bosonic field on the (metastable) vacuum at $\varphi^i =0$ it might decay or remain there in case the height and width of the barrier is large enough. We suppose that is the case here \cite{Landim:2016isc}. In such a case, making a background-perturbation split of the fields, they get expanded around the vacuum and the perturbations act as quantum fields in the interactions. The bosonic field $\varphi^+$, if trapped by a large enough barrier, can only decay by means of the Lagrangian, Eq. (\ref{eq:Lint}). Notice that since the false vacuum is at $\varphi^+=0$, the expanded Lagrangian coincides with the original one. Therefore, in the decay of the bosonic field $\varphi^+$ into fermions plus $\varphi^0$  there is no potential barriers, thus the dark energy decays into dark matter and other particles. This mechanism is the one responsible for the dark energy decay into cold (or even warm) dark matter and it is what we are going to pursue now.

From a cosmological point of view, the relevant interactions among the dark
particles are the decay $\varphi^+\rightarrow \varphi^0+\psi+\nu_d$
\cite{Landim:2016isc} and the annihilation of two scalars
into two fermions. The last process, however, gives a zero cross section after
expanding it in even powers of $p/m$ for a fermionic cold dark matter, while the previous (decay) process has already non relativistic contributions; generally speaking, it has more important contributions. The other annihilation processes belong to the hidden sector and do not play a major role in current observations.

\section{Model predictions}\label{modelpred}
\subsection{Background equations}

  The Boltzmann equation for a process $\alpha\rightarrow a+b+c$ is given by \cite{Kolb:1990vq}
\begin{eqnarray}\label{Bolteq1}
    \frac{\partial (a^3 n_\alpha)}{\partial t} &=&-a^3\int d\Pi_\alpha \, d\Pi_{a} \, d\Pi_{b} \, d\Pi_{c} \, (2\pi^4)\times {} \nonumber\\ 
    & {} & {} \times \delta^4(p_\alpha-p_a-p_b-p_c)  |\mathcal{M}|^2_{\alpha\rightarrow a+b+c}f_\alpha \quad, 
\end{eqnarray}
where $d\Pi_i\equiv \frac{1}{(2\pi)^3}\frac{d^3p_i}{2E_i}$ and $f_i=e^{-E_i/k_B
    T}$ and the $a = a(t)$ is the scale factor. We neglect the factors due
to  Bose condensation or Fermi degeneracy. 
The right-hand side of Eq.~(\ref{Bolteq1}) becomes
\begin{eqnarray}\label{Bolteq2}
    \int \! \! \! \! & {} & d\Pi_{\varphi^+} \, d\Pi_{\varphi^0} \, d\Pi_{\psi} \, d\Pi_{\nu} \, (2\pi^4) \delta^4(p_{\varphi^+}-p_{\varphi^0}-p_{\psi}-p_\nu) \times {} \nonumber \\
    & {} & {} \times |\mathcal{M}|^2e^{-E_{\varphi^+}/k_BT}= -\Gamma n_{\varphi^+}\quad,
\end{eqnarray}
where $\Gamma$ is the integral of the scattering amplitude, which in turn does
not depend on $P$. The number density is $n_{\varphi^+}=\int
e^{-E_{\varphi^+}/k_BT}\frac{d^3p_{\varphi^+}}{(2\pi)^3}$. Eqs.~(\ref{Bolteq1})
and (\ref{Bolteq2}) lead to the following equations for the particles in the
decay process
\begin{eqnarray}\label{Bolteq3}
\frac{\partial (a^3 n_\alpha)}{\partial t} &=&-\Gamma a^3 n_\alpha\quad,\\
\frac{\partial (a^3 n_{a,b,c})}{\partial t}&=&\Gamma a^3 n_{\alpha}\quad.
\end{eqnarray}

Once the field is at rest in
the minimum of the potential, from Eq.~(\ref{Bolteq3}) we see
that the term $a^3 n$ should be constant (in the absence of decay) to describe
the cosmological constant, therefore the energy density for a fluid with
equation of state $-1$ should be $\rho = a^3 m n$, that is, a
non-relativistic fluid that is not diluted as the universe
expands. 


The continuity equation for a cosmological fluid is obtained from the
definition $\rho_i \equiv \int \frac{d^3p_i}{(2\pi)^3}E_i f_i \approx m_i n_i$,
where the last equality holds for non-relativistic fluids. Hence, the continuity
equation for the $\varphi^+$ fluid is 
\begin{eqnarray}\label{continuity1}
    \dot{ \rho}_{\varphi^+}=-\Gamma \rho_{\varphi^+}\quad,
\end{eqnarray}
which has the usual exponential decay solution $\rho_{\varphi^+}\propto
e^{-\Gamma t}$. The decay rate can be seen as part of an effective equation of
state for $\varphi^+$, since
\begin{eqnarray}\label{continuity2}
    \dot{\rho}_{\varphi^+}+3H(1+w_{\mathrm{eff}}) \rho_{\varphi^+}=0\quad,
\end{eqnarray}
where $w_{\mathrm{eff}} =  -1 + \Gamma/3H$.
The second term gives rise to a kinetic contribution for the
dark energy.  

The other fluids ($\varphi^0$, $\psi$, and $\nu_d$) have similar continuity
equations, with the equations of state $w_{\varphi^+}=-1$, $w_\psi=0$ and either
$w_\nu=0$ or $w_\nu=1/3$. The two particles of the dark energy doublet and the
dark matter candidate are non-relativistic, which implies that the continuity
equations for the remaining fluids are
\begin{eqnarray}
    \dot{\rho}_{\varphi^0}&=& \mu_{\varphi^0} \Gamma \rho_{\varphi^+}\quad,\label{continuity31}\\
    \dot{\rho}_{\psi}+3H\rho_\psi &=& \mu_{\psi} \Gamma \rho_{\varphi^+}\quad,\label{continuity32}\\
    \dot{\rho}_{\nu}+3H(1+w_\nu)\rho_\nu&=&\bigl(1-\mu_{\varphi^0}-\mu_{\psi}\bigr) \Gamma \rho_{\varphi^+}\quad,\label{continuity33}
\end{eqnarray}
where in the last equation we have used the energy conservation
$E_\nu=m_{\varphi^+}-m_{\varphi^0}-m_\psi$, which is also evident from the
energy-momentum tensor conservation, and $\mu_{\varphi^0}$,
$\mu_{\psi}$ are the mass ratios $m_{\varphi^0}/m_{\varphi^+}$,
$m_{\psi}/m_{\varphi^+}$, respectively. The right-hand side of the
continuity equations above are a leading-order approximation since we are
considering non-relativistic fluids for $\varphi^+$, $\varphi^0$ and $ \psi$.

\subsection{Cosmological perturbations}

Once the equation of state parameters $w_i$ are constant for all fluids, their
sound speeds
are $c_{s,\, i}^2=\delta\mathcal{P}_i/\delta \rho_i= w_i$, where $\mathcal{P}_i$
is the pressure of the fluid `$i$'. 
The sound speed for a scalar
field is, in turn, $c_{s,\,\varphi}^2=1$ \cite{Gordon:2004ez}. Following the
definitions of \cite{Ma:1995ey}, in the synchronous gauge the energy
conservation leads to the following equations for the dark fluids
\begin{eqnarray}
\dot{\delta}_{\varphi^+}+6H\delta_{\varphi^+}&=&-\Gamma  \delta_{\varphi^+}\quad,\label{continuity41}\\
\dot{\delta}_{\varphi^0}+6H\delta_{\varphi^0}&=& \mu_{\varphi^0} \Gamma   \delta_{\varphi^+}\quad,\label{continuity42}\\
\dot{\delta}_{\psi}+\theta_\psi+\frac{\dot{h}}{2}&=& \mu_{\psi} \Gamma  \delta_{\varphi^+}\quad,\\
\dot{\delta}_{\nu}+(1+w_\nu)\left(\theta_\nu+\frac{\dot{h}}{2}\right)&=&\bigl(1-\mu_{\varphi^0}-\mu_{\psi} \bigr) \Gamma  \delta_{\varphi^+}\quad,\label{continuity44}\end{eqnarray}
where $\delta_i\equiv \delta \rho_i/\bar{\rho}_i$. The right-hand
sides of the equations above follow from Eqs.~(\ref{continuity2}--\ref{continuity33}).
Eq.~(\ref{continuity41}) has the solution $\delta_{\varphi^+}\propto
a^{-6}e^{-\Gamma t} $, in agreement with the fact that  dark energy does not
cluster on sub-horizon scales \cite{Duniya:2013eta}. Since the
$\varphi^+$ fluid is diluted in the universe faster than radiation, the
couplings in the right side
of Eqs.~(\ref{continuity42}--\ref{continuity44}) are negligible. As a result,
$\delta\varphi^0\propto a^{-6}$ and the continuity equation for the dark matter
perturbation is the same as in the uncoupled case. 
	
In order to get the interacting term in the momentum conservation equations, we
multiply the right-hand side of Eq.~(\ref{Bolteq2}) by
$p_{\varphi^+}/E_{\varphi^+}=p_{\varphi^+}/m_{\varphi^+}$ before integrating it.
The field velocity is defined as $v^i\equiv
\frac{1}{n}\int\,d^3p{\frac{p\hat{p}^i}{E}e^{-E/T}}$, thus the Navier-Stokes
equation in momentum space for $\varphi^+$ is
\begin{eqnarray}
    k^2\delta_{\varphi^+}&=\theta_{\varphi^+}\Gamma\quad, \label{continuity51}
\end{eqnarray}
where $\theta\equiv ik_j v^j$.  The field velocity for $\varphi^+$ is also
negligible because the left-hand side of Eq.~(\ref{continuity51}) goes to zero.
Thus the momentum transfer is irrelevant and the Navier-Stokes equation for
dark matter is the usual one from the $\Lambda$CDM model. Therefore, the decay
process changes only the background equations. 

\subsection{Comparison with observations}
	
\begin{figure}
    \centering
    \includegraphics[width=0.6\columnwidth]{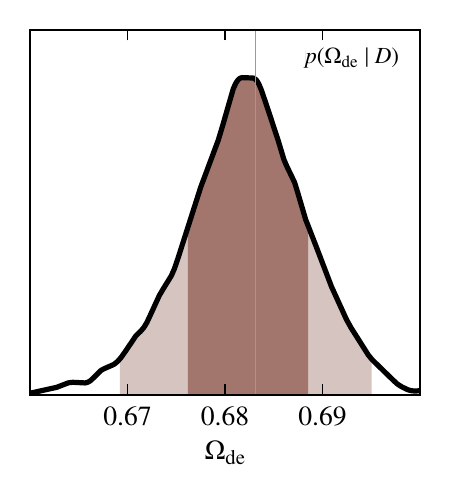}
    \caption{Marginalized posterior probability $p$ of the derived parameter
        $\Omega_{\de} \equiv \Omega_{\varphi^+} + \Omega_{\varphi^0}$ given the
        data $D$ from standard cosmic probes.
        The shaded areas under the curve mark the $1\sigma$ and $2\sigma$
        confidence levels; the thin grey line marks the parameter value at the
        best-fit (bf) point,
        $\Omega_{\de}^{\mathrm{(bf)}} = 0.68313$.}
    \label{fig:Ode_dist}
\end{figure}

From the observational point of view,
the two scalar fields $\varphi^+$ and $\varphi^0$ have $w_i = -1$
and effectively behave like one ``dark energy'' fluid.
The same happens with the two particles in the dark matter doublet in the case
that the dark neutrino is non-relativistic ($w_{\nu} = 0$). For this doublet,
the background equations (\ref{continuity32}) and (\ref{continuity33}) can be
combined into
\begin{eqnarray}
     \dot{\rho}_{\dm}+3H\rho_{\dm}&=&\bigl(1-\mu_{\varphi^0}\bigr) \Gamma \rho_{\varphi^+}\quad\label{continuity33together}.
\end{eqnarray}
It is then straightforward to solve numerically the background cosmology in
terms of the scale factor,
\begin{eqnarray}
    \frac{d\rho_{\phi^+}}{da} &=& - \frac{\Gamma}{a H} \rho_{\varphi^+}, \label{contnum}\\
    \frac{d\rho_{\varphi^0}}{da} &=& \mu_{\varphi^0} \frac{\Gamma}{a H} \rho_{\varphi^+}, \\
    \frac{d\rho_{\dm}}{da} &=& - \frac{3}{a} \rho_{\dm} + \bigl(1 - \mu_{\varphi^0} \bigr) \frac{\Gamma}{a H} \rho_{\varphi^+},
\end{eqnarray}
backwards in time with the current densities as ``initial'' conditions,
together with the usual equations for the standard model fluids. Rewriting the equations in terms of the scale factor eases the numerics.
A degeneracy between $\Gamma$ and the density of $\varphi^{+}$ is evident from
these equations.
The two parameters always appear multiplied.
Writing them in terms of a new ``density'' $\rho_{\Gamma} \equiv \Gamma
\rho_{\varphi^{+}}$ gives
\begin{eqnarray}
    \frac{d\rho_{\Gamma}}{da} &=& - \frac{\Gamma}{a H} \rho_{\Gamma}, \\
    \frac{d\rho_{\varphi^0}}{da} &=& \frac{\mu_{\varphi^0}}{a H} \rho_{\Gamma}, \\
    \frac{d\rho_{\dm}}{da} &=& - \frac{3}{a} \rho_{\dm} + \frac{1 - \mu_{\varphi^0}}{a H} \rho_{\Gamma},
\end{eqnarray}
which partially amends the problem.
However, we fail to obtain constraints on the relevant parameters
of the decay (now $\Gamma$, $\Omega_{\Gamma}$, $\Omega_{\dm}$,
$\mu_{\varphi^0}$, and $\Omega_{\varphi^0}$ determined from the flatness
condition on the sum of the density parameters) when we compare the predicted
evolution with the standard cosmic probes using Markov Chain Monte Carlo 
(MCMC) simulations.
Despite this, the derived parameter $\Omega_{\de} \equiv \Omega_{\varphi^+} +
\Omega_{\varphi^0}$, with $\Omega_{\varphi^+} = \Omega_{\Gamma}/\Gamma$ is
verified to mimic the standard model`s dark energy with
$\Omega_{\de} = 0.68183^{+0.00668}_{-0.00564}$ at $1 \sigma$ confidence level
(see Fig.~\ref{fig:Ode_dist}).
For this analysis, we employed observational data from the
Planck Cosmic Microwave Background (CMB) distance priors
\cite{huang_distance_2015}, the Joint Light-curve Analysis (JLA) of type-Ia
supernovae \cite{betoule_improved_2014}, Baryon Acoustic Oscillation (BAO) from various surveys
\cite{beutler_6df_2012,ross_clustering_2015,anderson_clustering_2014,fontribera_2014},
$H(z)$ from cosmic clocks \cite[and references therein]{moresco_6_2016},
and the local value of the Hubble constant \cite{riess_2016}.
Because the parameters $\Omega_{\Gamma}$, $\Gamma$ and $\mu_{\varphi^0}$ are
expected to be small, we adopted conservative flat priors restricting them to the
interval $\left[0, 0.5 \right]$. 

The difficulties discourage any further attempt to constrain the parameters of
this model in the case $w_{\nu} = 1/3$, which adds two more parameters,
$\Omega_{\nu}$ and $\mu_{\psi}$ (the dark matter doublet cannot be described as
a single fluid anymore), potentially making the degeneracy even more serious.

\section{Conclusions}\label{conclusion}

In this paper we investigated the interactions among the dark particles in the
dark $SU(2)_R$ model from a cosmological point of view. The most relevant
interaction is the decay of one particle in the dark energy doublet into the
other three particles in the dark energy and dark matter doublet. This process
consists of a new form of interacting dark energy and it changes only the
background equations. Although the comparison with data constrained very well
the dark energy density parameter today, defined as the sum of the density
parameters of $\varphi^+$ and $\varphi^0$, the other free parameters in the
process (decay rate and masses of the particles) are not constrained mainly due
to the strong degeneracy between the decay rate and the density of the
progenitor ($\varphi^+$). 

\begin{acknowledgements}

This work is supported by CAPES, CNPq and FAPESP. RL would like to thank Orfeu Bertolami and the Departamento de F\'isica e Astronomia, Faculdade de Ci\^encias, Universidade do Porto for hosting him while the work was in progress. This work has made use of the computing facilities of the Laboratory of Astroinformatics (IAG/USP, NAT/Unicsul), whose purchase was made possible by the Brazilian agency FAPESP (grant 2009/54006-4) and the INCT-A.

\end{acknowledgements}

\bibliographystyle{unsrt}
\bibliography{references}
\end{document}